\documentclass[amssymb,prl,aps,twocolumn,amsmath,showpacs]{revtex4} %,preprint

\usepackage{graphicx}

\begin{document}

\title{Experimental Test of the Dynamical Coulomb Blockade Theory for Short Coherent Conductors}
\author{C. Altimiras}
\affiliation{Phynano team, Laboratoire de Photonique et de
Nanostructures (LPN) - CNRS, route de Nozay, 91460 Marcoussis,
France}
\author{U. Gennser}
\affiliation{Phynano team, Laboratoire de Photonique et de
Nanostructures (LPN) - CNRS, route de Nozay, 91460 Marcoussis,
France}
\author{A. Cavanna}
\affiliation{Phynano team, Laboratoire de Photonique et de
Nanostructures (LPN) - CNRS, route de Nozay, 91460 Marcoussis,
France}
\author{D. Mailly}
\affiliation{Phynano team, Laboratoire de Photonique et de
Nanostructures (LPN) - CNRS, route de Nozay, 91460 Marcoussis,
France}
\author{F. Pierre}
\email[Corresponding author: ]{frederic.pierre@lpn.cnrs.fr}
\affiliation{Phynano team, Laboratoire de Photonique et de
Nanostructures (LPN) - CNRS, route de Nozay, 91460 Marcoussis,
France}

\date{\today}

\begin{abstract}
We observed the recently predicted quantum suppression of dynamical
Coulomb blockade on short coherent conductors by measuring the
conductance of a quantum point contact embedded in a tunable on-chip
circuit. Taking advantage of the circuit modularity we measured most
parameters used by the theory. This allowed us to perform a reliable
and quantitative experimental test of the theory. Dynamical Coulomb
blockade corrections, probed up to the second conductance plateau of
the quantum point contact, are found to be accurately normalized by
the same Fano factor as quantum shot noise, in excellent agreement
with the theoretical predictions.
\end{abstract}


\maketitle

A tunnel junction exhibits a drop of its conductance at low voltages
and temperatures when it is embedded in a resistive circuit, in
violation of the classical impedances composition laws. This quantum
phenomenon, known as dynamical Coulomb blockade (DCB), results from
the excitation of the circuit's electromagnetic modes by the current
pulses associated with tunnel events. The theory is well understood
and verified experimentally for tunnel junctions
\cite{SCT,DEVetal1990_GIRVINetal1990,CLELANDetal1992_HOLSTetal1994,JOYEZetal1998,PIERREetal2001},
but it is only recently that it has been extended to short coherent
conductors \cite{GZ2001_LEVYYEYATIetal2001}. The strong recent
prediction is that DCB corrections are simply reduced by the
\textit{same} normalization factor as quantum shot noise, as a
consequence of electron flow regulation by the Pauli exclusion
principle \cite{MARTINLANDAUER1992}. The aim of this work is to
perform an accurate experimental test of the DCB theory for coherent
conductors and thereby to provide solid grounds to our knowledge of
impedances composition laws in mesoscopic circuits.

A powerful description of coherent conductors in absence of
interactions is provided by the scattering approach, which
encapsulates the complexity of transport mechanisms into the set
$\{\tau_n\}$ of transmission probabilities across the conduction
channels indexed by $n$. In short conductors, the energy dependence
of $\{\tau_n\}$ can be neglected provided that $h/\tau_{dwell}\gg
k_BT,eV_{SD}$, where $\tau_{dwell}$ is the dwell time in the
conductor, $T$ the temperature and $V_{SD}$ the applied voltage
\cite{BUTTIKERLANDAUER1986}. The conductance then reads
$G=G_Q\sum_{n}\tau_n$, with $G_Q=2e^2/h$ the conductance quantum;
and the current shot noise at zero temperature is $S_I=2eIF$, where
$2eI$ is the Poissonian noise and
$F=\sum_{n}\tau_n(1-\tau_n)/\sum_{n}\tau_n$ is the Fano factor. More
generally, the full counting statistics of charge transfers can be
formulated with $\{\tau_n\}$ \cite{BLANTERBUTTIKER2000}. How is this
picture modified by Coulomb interaction? First, the low energy
excitations are transformed from electrons to Fermion quasiparticles
of finite lifetime which thereby limits the coherent extent of
conductors \cite{PINESNOZIERE}. Second, Coulomb interaction couples
a coherent conductor to the circuit in which it is embedded, which
results in the DCB. In practice, DCB corrections reduce the
transmission probabilities at low energies. The theory of DCB has
first been worked out for small tunnel junctions of resistance large
compared to the resistance quantum $R_K=h/e^2\simeq 25.8$~k$\Omega$
and embedded in macroscopic linear circuits characterized by a
frequency dependent impedance $Z_{env}(\nu)$
\cite{SCT,DEVetal1990_GIRVINetal1990}. The theory has been found in
excellent agreement with experiments
\cite{CLELANDetal1992_HOLSTetal1994}, and more recently extended to
low impedance \cite{JOYEZetal1998} and long \cite{PIERREetal2001}
tunnel junctions. From a theoretical standpoint, tunnel junctions
are easy to deal with since they can be treated perturbatively. The
generalized DCB theory to short coherent conductors, whose
transmission probabilities can take any value between 0 and 1,
assumes instead that quantum fluctuations are small. This hypothesis
limits its validity to low environmental impedance
$\mathrm{Re}[Z_{env}(\nu)]\ll R_K$. The striking prediction is that
the amplitude of DCB corrections to the conductance of coherent
conductors is reduced relative to tunnel junctions by the same Fano
factor as quantum shot noise \cite{GZ2001_LEVYYEYATIetal2001}.
Further theoretical investigations concluded that a similar relation
holds more generally between the Coulomb corrections to the $n$th
cumulant of current fluctuations and the $(n+1)$-th cumulant
\cite{GALAKTIONOVetal2003_KINDERMANNetal2003_KINDERMANNetal2004_SAFI2004}.
Experimentally, a pioneer work performed on an atomic contact showed
that DCB corrections are strongly reduced when the transmission
probability approaches 1, in qualitative agreement with the theory
\cite{CRONetal2001}. However, as pointed out by the authors of
\cite{CRONetal2001}: ``it (was) not possible to conclude whether or
not (the theory) is quantitatively correct". Indeed, at large
transmissions, relatively large universal conductance fluctuations
were superimposed on the DCB signal whereas, in the tunnel regime,
the set of transmission probabilities could not be extracted
reliably due to significant DCB corrections. Up to now a
quantitative test of the dynamical Coulomb blockade theory for a
coherent conductor was missing. The present experiment fills this
gap.

\begin{figure}[tb]
\includegraphics[width=3.3in]{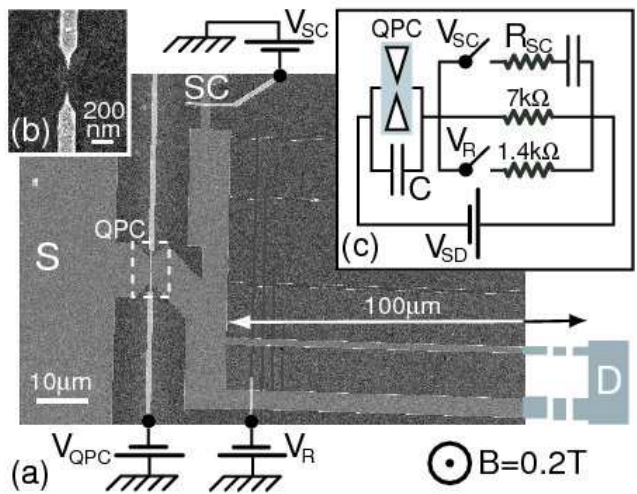} \caption{(a) E-beam
micrograph of the sample tailored in a GaAs/Ga(Al)As heterojunction.
The 2DEG is patterned by chemical etching, etched areas are darker.
Active top metal gates are colorized in the lighter grey. Electrode
labels S, D and SC stand respectively for Source, Drain and Short
Circuit. (b) Magnified view of the metallic split gate used to tune
the QPC. (c) Schematic representation of the sample.} \label{Fig1}
\end{figure}

In this experiment, we have measured the variations in the
resistance of a quantum point contact (QPC) realized in a 2D
electron gas (2DEG) while changing the adjustable on-chip circuit in
which it is embedded. The conduction channels of a QPC are directly
related to the 1D sub-bands quantized by the transverse confinement
\cite{GLAZMANetal1988}. By reducing the confinement with voltage
biased top gates, the transmission probabilities of the conduction
channels are increased continuously and, for adequate geometries
\cite{BUTTIKER1990}, one channel at a time. Consequently, the QPC's
conductance $G_{QPC}=(n+\tau_{n+1})G_Q$ corresponds to $n$ channels
fully transmitted and one channel of transmission probability
$\tau_{n+1}$. The knowledge of the transmission probabilities
combined with the ability to change them continuously make of a QPC
a powerful test-bed for short coherent conductors
\cite{REZNIKOVetal1995_KUMARetal1996}. As described later, we can
change in-situ the circuit surrounding the QPC using voltage biased
metallic top gates to deplete the 2DEG underneath. It is by
monitoring the QPC's resistance as a function of the circuit
impedance that we can extract accurately the amplitude of DCB
corrections.

The measured sample, shown in Fig.~1, was realized in a
GaAs/Ga(Al)As heterojunction. The 2DEG is 94~nm deep, of density
$2.5~10^{15}~\mathrm{m}^{-2}$, Fermi energy $100~$K and mobility
$55~\mathrm{m}^2V^{-1}s^{-1}$. The sample was patterned using e-beam
lithography followed by chemical etching of the heterojunction and
by deposition of metallic gates at the surface. The QPC is formed in
the 2DEG by applying a negative voltage $V_{QPC}$ to the metallic
split gates shown in Fig.~1(b). Two stripes of width $1.4~\mu$m and
$3.8~\mu$m \cite{noteWEFF}, and of length $100~\mu$m, much longer
than the electron phase coherence length $L_\phi \sim 10~\mu$m, were
patterned in the 2DEG by chemical etching to form an on-chip
resistance in series with the QPC. Measurements were performed in a
dilution refrigerator of base temperature T$=40$~mK. All measurement
lines were filtered by commercial $\pi$-filters at the top of the
cryostat. At low temperature, the lines were carefully filtered and
thermalized by arranging them as 1~m long resistive twisted pairs
($300~\Omega /$m) inserted inside 260~$\mu$m inner diameter CuNi
tubes tightly wrapped around a copper plate screwed to the mixing
chamber. The sample was further protected from spurious high energy
photons by two shields, both at base temperature. Conductance
measurements were performed using standard lock-in techniques at
excitation frequencies below 100~Hz. The sample was current biased
by a voltage source in series with a $10~$M$\Omega$ or
$100~$M$\Omega$ polarization resistance at room temperature.
Voltages across the sample were measured using low noise room
temperature amplifiers. The source (S)-drain (D) voltage was kept
smaller than $k_BT/e$ to avoid heating. We applied a small
perpendicular magnetic field B$=0.2$~T \cite{noteB} to minimize
non-ideal behaviors of the QPC such as sharp energy dependence of
the transmissions resulting from Fabry-P\'{e}rot resonances with
nearby defects, and imperfect transmissions across ``open" channels.

In our experiment the QPC is embedded in an electromagnetic
environment schematically represented as a R//C circuit in
Fig.~1(c). The parallel capacitance ($C$) is the geometrical
capacitance between the source electrode (S) and the vertical near
rectangular conductor on the right side of the QPC. If the short
circuit electrode (SC) is disconnected ($V_{SC}<-0.3$~V), the
on-chip series resistance can take two values
$R_S=1.2~\mathrm{k}\Omega$ and $7~\mathrm{k}\Omega$ depending on
whether the wider 2DEG stripe is, respectively, connected ($V_R=0$)
or disconnected ($V_R=-0.35$~V), using the metal gate voltage $V_R$
as a switch. If the SC electrode is connected ($V_{SC}\simeq0$), it
acts as a low impedance ($R_{SC}$) high frequency path to ground in
parallel with $R_S$. Note that DCB reduces the DC conductance of a
coherent conductor but that these DCB corrections depend on the
impedance of the electromagnetic environment at high frequencies,
typically $\nu\sim k_BT/h\in [0.8,4]$~GHz for $T\in [40,200]$~mK.
Consequently, while the SC electrode is connected at room
temperature to a high input impedance voltage amplifier, at high
frequencies the environment impedance is expected to be reduced to
the on-chip resistance of the SC electrode plus, approximately, the
vacuum impedance $377~\Omega$ due to antenna effects on length
scales larger than a fourth of the electromagnetic wavelength. This
is symbolized in Fig.~1(c) by a high frequency impedance $R_{SC}$ in
series with a capacitor that acts as a high frequency short circuit.

The experiment was performed as follows: \textit{i)} We first
selected a series resistance $R_S=1.2~\mathrm{k}\Omega$ or
$7~\mathrm{k}\Omega$ with $V_R$. \textit{ii)} With the short circuit
electrode (SC) connected ($V_{SC}\simeq0$), we tuned the QPC with
$V_{QPC}$. In this configuration the DCB corrections are minimum
because the series resistance $R_S$ is shorted at high frequency by
$R_{SC}$. Since the SC electrode is disconnected from ground at the
near DC frequencies applied to measure the sample, it could be used
to measure separately the QPC and the series resistances.
\textit{iii)} We then disconnected the SC electrode by applying a
negative voltage $V_{SC}$, therefore increasing the high frequency
circuit impedance and consequently the DCB corrections. By
simultaneously measuring the variations of the source (S)-drain (D)
resistance, which is the sum of the QPC and the series resistance,
we can extract the amplitude of DCB corrections.

\begin{figure}[tb]
\includegraphics[width=3in]{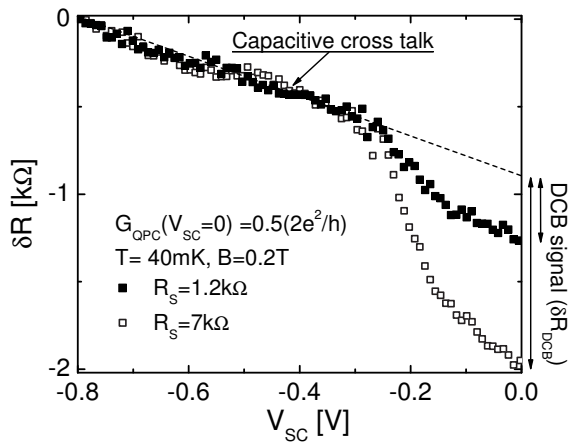} \caption{Resistance variation $\delta$R
of the QPC in series with the on-chip resistance
$R_S=1.2~\mathrm{k}\Omega$ ($\blacksquare$) or
$R_S=7~\mathrm{k}\Omega$ ($\square$) plotted versus the voltage
$V_{SC}$ that controls the high frequency short circuit (SC) switch
(see Fig.~1(c)). For $V_{SC}<-0.3$~V the SC switch is open and
$\delta$R exhibits a linear dependence with $V_{SC}$ due to the
direct capacitive cross talk with the QPC. The DCB signal $\delta
R_{DCB}$ is the difference between the resistance measured at
$V_{SC}\simeq0$ (SC switch closed) and the resistance measured for
an open SC switch taking into account the linear capacitive
contribution (dashed line).} \label{Fig2}
\end{figure}

Figure~2 shows $\delta R$, the resistance variation of the QPC plus
the series resistance from their values at $V_{SC}=-0.8$~V, plotted
versus $V_{SC}$ at $G_{QPC}(V_{SC}=0)=0.5G_Q$, T$=40$~mK and
B$=0.2$~T for $R_S=1.2~\mathrm{k}\Omega$ and $7~\mathrm{k}\Omega$.
The dependence of $\delta R$ with $V_{SC}$ results from two
contributions: \textit{i)} At $V_{SC}<-0.3$~V the SC electrode is
disconnected and $\delta R$ is a linear function of $V_{SC}$ with a
negative slope that does not depend on $R_S$. This is a consequence
of the capacitive cross talk between the metal gate controlled by
$V_{SC}$ and the QPC. We have checked (data not shown) that this
slope, which is a non monotonous function of $V_{QPC}$, is
proportional to the derivative of the QPC's resistance with
$V_{QPC}$. The normalization factor $\simeq 10^{-3}$ is in rough
quantitative agreement with the sample geometry. \textit{ii)} For
$V_{SC}>-0.3$~V we observe, on top of the linear capacitive cross
talk, a sudden drop when $V_{SC}$ increases. We attribute this
resistance drop, written hereafter $\delta R_{DCB}$, to the
reduction of DCB corrections as the parallel high frequency short
circuit electrode SC gets connected. As expected, $\delta R_{DCB}$
is larger in the more resistive environment
$R_S=7~\mathrm{k}\Omega$. In the following we extract $\delta
R_{DCB}$ by measuring the QPC in series with $R_S$ successively at
$V_{SC}=-0.1$~V (SC electrode connected) and $V_{SC}=-0.33$~V (SC
electrode disconnected). We then subtract the capacitive cross talk
contribution obtained from $\delta R(V_{SC}=-0.56$~V$)-\delta
R(V_{SC}=-0.33$~V$)$.

\begin{figure}[tb]
\includegraphics[width=3in]{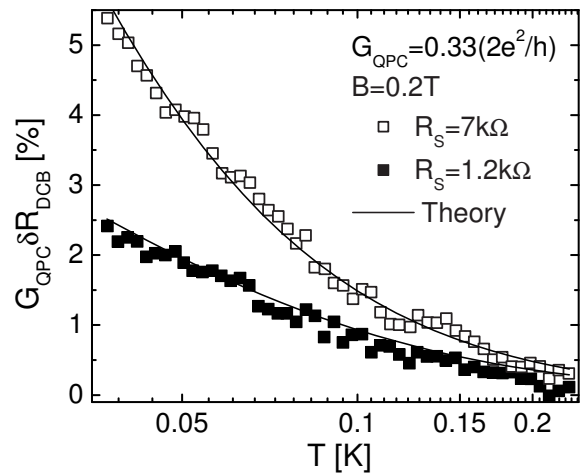} \caption{Measured temperature
dependence of relative DCB corrections $G_{QPC}\delta R_{DCB}$ at
$G_{QPC}=0.33G_Q$ and B$=0.2$~T, for $R_S=1.2$~k$\Omega$
($\blacksquare$) and $R_S=7$~k$\Omega$ ($\square$). Predictions of
the DCB theory are shown as continuous lines. The only fit parameter
in the calculation is the high frequency residual resistance of the
short circuit path $R_{SC}=1$~k$\Omega$ (see text).} \label{Fig3}
\end{figure}

Figure~3 shows as symbols the measured temperature dependence of the
DCB signal at $G_{QPC}(V_{SC}=-0.1~\mathrm{V})=0.33G_Q$ for
$R_S=1.2$~k$\Omega$ and $7$~k$\Omega$. The continuous lines are
predictions of the DCB theory for tunnel junctions \cite{SCT},
normalized by the one-channel Fano factor $F=1-G_{QPC}/G_Q \simeq
0.67$. The schematic R//C circuit modeling the QPC's electromagnetic
environment is shown in Fig.~1(c). The real part of its impedance
plugged into the theory reads $\mathrm{Re}[Z_{env}(\nu)]=R/(1+(2\pi
R C \nu)^2)$. The calculated $\delta R_{DCB}$ is the difference in
the amplitude of DCB corrections for open and closed short circuit
switch. The corresponding circuit resistance $R$ is, respectively,
$R=R_S$ and $R=1/(1/R_S+1/R_{SC})$. The only fit parameter in our
calculation is the SC high frequency impedance that we fixed at
$R_{SC}=1$~k$\Omega$, in agreement with the sum of the on-chip SC
resistance estimated from the geometry to 600$\pm100$~$\Omega$ and
the vacuum impedance 377~$\Omega$. Other parameters plugged into the
DCB calculation are the measured series resistances
$R_S=1.2$~k$\Omega$ or $R_S=7$~k$\Omega$ and the geometrical
capacitance $C=30$~fF estimated numerically with an accuracy of
$\pm5$~fF \cite{noteCNUMERICAL}. The very good agreement between
data and theoretical predictions provides a strong support to our
interpretation and allows us to now compare the measured dependence
of DCB on transmission probabilities with the predicted Fano
reduction factor.

\begin{figure}[tb]
\includegraphics[width=3.3in]{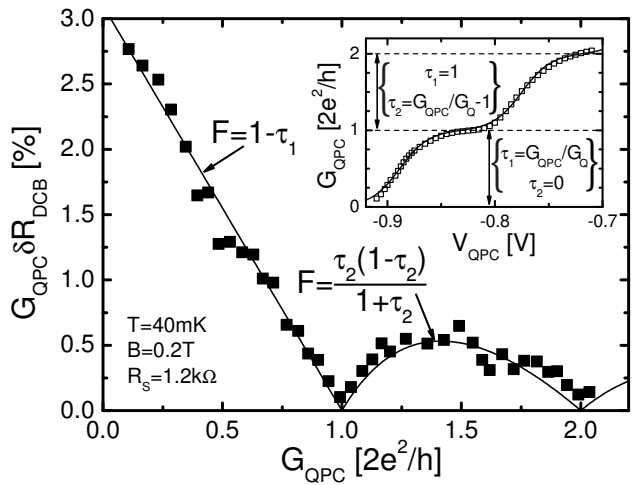} \caption{Measured
relative DCB corrections ($\blacksquare$) plotted versus the QPC
conductance. The continuous line is the predicted reduction by the
Fano factor $F=\sum_{n}\tau_n(1-\tau_n)/\sum_{n}\tau_n$. Inset:
Measured QPC conductance ($\square$) plotted versus the split gate
bias voltage $V_{QPC}$. The best fit using a quadratic confinement
potential \cite{BUTTIKER1990} is shown as a continuous line.}
\label{Fig4}
\end{figure}

To test the generalized dynamical Coulomb blockade theory, we
measured the relative amplitude of DCB corrections versus the QPC
conductance at 40~mK and for $R_S=1.2$~k$\Omega$ \cite{noteUCF} (see
Fig.~4). The predictions depend on the set $\{\tau_n\}$, it is
therefore crucial to extract accurately the transmissions
probabilities of the QPC. The inset in Fig.~4 shows the QPC
conductance up to $2G_Q$ versus the split gate voltage $V_{QPC}$,
measured at T=40~mK and B=0.2~T. We subtracted 350~$\Omega$ from the
data to account for the residual DC series resistance by adjusting
the first three plateaus on multiples of the conductance quantum
\cite{noteGQPC}. From the maximum deviation between our data and the
best fit (continuous line in inset of Fig.~4) using B\"{u}ttiker's
model of QPCs \cite{BUTTIKER1990}, we estimate our accuracy on the
transmission probabilities
$\{\tau_1=\min[1,G_{QPC}/G_Q],\tau_2=\max[0,G_{QPC}/G_Q-1]\}$ to be
better than 0.05. The continuous line in Fig.~4 shows the relative
amplitude of DCB as predicted by theory
\cite{GZ2001_LEVYYEYATIetal2001}. We observe an excellent
quantitative agreement between the data and the Fano factor
$F=(\tau_1(1-\tau_1)+\tau_2(1-\tau_2))/(\tau_1+\tau_2)$ that
controls quantum shot noise \cite{REZNIKOVetal1995_KUMARetal1996}.

To conclude, we have performed a quantitative experimental test of
the generalization of dynamical Coulomb blockade theory to short
coherent conductors embedded in low impedance circuits. We find
dynamical Coulomb blockade corrections that are reduced in amplitude
by the same Fano factor as quantum shot noise, in quantitative
agreement with the predictions. This result is not only important
within the fundamental field of quantum electrodynamics in
mesoscopic circuits. It also provides solid grounds to engineer
complex devices with coherent conductors and to use dynamical
Coulomb blockade as a tool to probe the transport mechanisms. For
this purpose DCB has the advantage on shot noise that the signal
increases when the probed energies decrease.

The authors gratefully acknowledge inspiring discussions and
suggestions by D.~Est\`{e}ve, P.~Joyez, H.~Pothier and C.~Urbina. We
also thank M.H.~Devoret, F.~Portier and B.~Reulet for stimulating
discussions and G.~Faini, R.~Giraud and Y.~Jin for permanent
assistance. This work was supported by the ANR (ANR-05-NANO-039-03)
and NanoSci-ERA (ANR-06-NSCI-001).


\begin{references}
\bibitem{noteWEFF} The 2DEG width is about 700~nm narrower due to lateral depletion near the edges.
\bibitem{noteB} Shubnikov-de Haas oscillations of $R_S$(B) are negligible up to the applied magnetic field B=0.2~T.
\bibitem{noteGQPC} The corresponding reduction of $G_{QPC}$ is at most 2.5\% and 5\% below, respectively, the
first and second plateau.
\end{references}
\end{document}